\documentclass[10pt,letterpaper]{article}
\usepackage[top=0.85in,left=1.25in,footskip=0.75in]{geometry}

\usepackage{helvet}
\usepackage{booktabs}

\usepackage{amsmath,amssymb}

\usepackage{changepage}

\usepackage{textcomp,marvosym}

\usepackage{cite}

\usepackage{nameref,hyperref}

\usepackage[right]{lineno}

\usepackage[nopatch=eqnum]{microtype}
\DisableLigatures[f]{encoding = *, family = * }

\usepackage[table]{xcolor}

\usepackage{array}

\newcolumntype{+}{!{\vrule width 2pt}}

\newlength\savedwidth

\raggedright
\usepackage[aboveskip=1pt,labelfont=bf,labelsep=period,justification=raggedright,singlelinecheck=off]{caption}

\makeatletter
\renewcommand{\@biblabel}[1]{\quad#1.}
\makeatother

\usepackage{lastpage,fancyhdr,graphicx}
\usepackage{epstopdf}
\fancyheadoffset[L]{2.25in}
\fancyfootoffset[L]{2.25in}

\begin{document}
\vspace*{0.2in}

\begin{flushleft}
{\Large
\textbf\newline{Shifting norms in scholarly publications: \\ trends in readability, objectivity, authorship, and AI use} 
}
\newline
\\
P\'{a}draig Cunningham\textsuperscript{1*},
Padhraic Smyth\textsuperscript{2},
Barry Smyth\textsuperscript{1*}
\\
\bigskip
\textbf{1} School of Computer Science, University College Dublin, Ireland\\
\textbf{2} Department of Computer Science, University of California, Irvine, USA
\bigskip

%
%


%
* \{padraig.cunningham,barry.smyth\}@ucd.ie

\end{flushleft}
\section*{Abstract}
Academic and scientific publishing practices have changed significantly in recent years. This paper presents an analysis of 17 million research papers published since 2000 to explore changes in authorship and content practices. It shows a clear trend towards more authors, more references and longer abstracts. While increased authorship has been reported elsewhere, the present analysis shows that it is pervasive across many major fields of study. We also identify a decline in author productivity which suggests that `gift' authorship (the inclusion of authors who have not contributed significantly to a work) may be a significant factor. We further report on a tendency for authors to use more hyperbole, perhaps exaggerating their contributions to compete for the limited attention of reviewers, and often at the expense of readability. This has been especially acute since 2023, as AI has been increasingly used across many fields of study, but particularly in fields such as Computer Science, Engineering and Business. In summary, many of these changes are causes of significant concern. Increased authorship counts and gift authorship have the potential to distort impact metrics such as field-weighted citation impact and $h$-index, while increased AI usage may compromise readability and objectivity.


\section{Introduction}


As the world of science has been changing \cite{10.1145/3097983.3098016}, so too has academic publishing \cite{doi:10.1128/mbio.02515-24}. Over the past 25 years new fields and specialties have emerged. There have been changes in how science is funded, and the roles of academia and industry have been evolving. The number of scientific journals and venues has been increasing, alongside new publishing models. This has been driven in part by the Open Access (OA) movement and the need for shortened review times to make research more rapidly and freely available to the public, but perhaps at the expense of impact and quality \cite{cunningham2024analysis}.

Indeed, there is growing concern that pressures on researchers to publish may have a negative impact on publication behavior \cite{osborne2019authorship,mills2021problematizing,severin2021overburdening,hanson2024strain}. Hanson \emph{et al.} \cite{hanson2024strain} show that there has been a significant growth in research papers published in recent years, with the overall number of papers available in 2022 being 47$\%$ higher than in 2016. They argue that this is due to the unfortunate dynamic that exists between the three main parties (funding agencies, researchers and publishers), all of whom are motivated by volume rather than quality. 


In this paper we present results of an analysis of a Semantic Scholar \footnote{\url{semanticscholar.org}} dataset containing 17M papers across 23 fields of study. We show that publication practices across many fields have changed significantly over the period 2000 - 2024, but especially since the arrival of modern AI tools in 2023. Our analysis focuses on two aspects of scientific publication practices: (i) authorship practices and productivity; and (ii) content practices, such as the use of  AI-generated content and hyperbole and the subsequent impact on objectivity and readability. 

We show that the number of papers being published per year has been steadily increasing over time, alongside the average number of authors per paper. These rates of increase vary by field of study. However, when we adjust for the relative contribution of multiple authors on papers -- assuming a typical author will contribute more to a two-author paper than to a four-author paper -- we find a decline in indicators of author productivity. Different fields exhibit different degrees of productivity decline with, for example, Philosophy, Mathematics and Computer Science presenting with higher productivity levels, when we control for authorship.

Our results also show a gradual but sustained decrease in objectivity and readability over the last 25 years. There has been a steady increase in the use of hyperbole (hype terms) across all 23 fields of study, and a corresponding decline in readability. Moreover, there is evidence of a strong AI signal in the content of abstracts since the emergence of AI tools and commercial large language models (LLMs) in late 2022. This AI signal is associated with  the decline in readability and a marked acceleration in the use of hype terms. We find these effects vary significantly across the various fields of study with the fields of Computer Science and Engineering exhibiting an especially strong AI signal.


These results are important for several reasons. The fact that author productivity is declining in the face of rising publication volume and increased author counts, is surprising. As we shall discuss, it may reflect the increased cost of multi-author collaborations, or it may be due to instances of gift authorship, or a combination of both. Moreover, while the use of AI is likely to continue, the strong association with a loss of objectivity (increased use of hype terms) and declining readability is a notable secondary effect. The scientific publishing world needs to adapt to these changing authorship practices. As open access models have influenced publication impact and quality \cite{cunningham2024analysis},
it is also important to consider ongoing shifts in authorship, research productivity, and the use of AI—particularly in relation to how research quality is evaluated through metrics such as the $h$-index and field-weighted citation scores.

\section{Related research}
This work considers how authorship and content practices have been changing in scholarly publications over the last several decades. It is not the first attempt to look at such issues, but it is distinguished in part by the scale of the analysis offered, in terms of the number of papers analysed and the variety of research fields considered. In this section we review related work in   areas of overlap with the present work including changing author counts, the use of AI tools, and changes in language use and readability.

\subsection{Changing author counts}
\label{sec:changineauthorcounts}

A number of researchers \cite{jakab2024many,cunningham2024analysis} have recently highlighted an increase over time in the numbers of authors on papers. While some drivers of this increase may not be a cause for concern, such as an increase in interdisciplinary/collaborative research, there is concern that this increase may  be an indicator of some poor research practices \cite{hosseini2022ethical}. 

One particular concern is what is often termed `gift' authorship \cite{hosseini2022ethical}. This could be the inclusion of senior colleagues as author on a paper without sufficient contribution to the work (also called honorary authorship) or the unwarranted sharing of authorship among junior authors to help progress their careers. It is important to recognise that gift (honorary) authorship is not new; a study from 1998 by Flanagin \emph{et al.} found that it was already prevalent in medical publications at that time \cite{flanagin1998prevalence}. Indeed Reisig \emph{et al.} \cite{reisig2020assessing} contend that authorship fraud is the most common type of research fraud in US universities. 

Brand \emph{et al.} report that between 1930 and 1970 the average number of authors on scientific papers remained constant at roughly two authors\cite{brand2015beyond}. They argue that the increase observed since the 1970s is likely due in part to some poor authorship practices, and they argue for a clear framework for recognizing author contribution in order to address this issue. This framework called CRediT 
\href{https://credit.niso.org}{credit.niso.org} provides a taxonomy of 14 roles against which author contributions on a paper can be recorded. The expectation being that a detailed account of author contribution will address the issue. 

Zauner \emph{et al.} write about authorship as a commodity that can be `gifted' or purchased
 \cite{zauner2018we}. They support a proposal for a set of four criteria that needs to be met for authorship. There needs to be a substantial contribution to the work itself, a contribution on that drafting of the paper, approval of the published version and accountability for all aspects of the work. Authors should contribute on all four aspects.

\subsection{Use of AI}
The use of AI in scientific writing has been researched extensively since the arrival of ChatGPT in late 2022. How to detect AI-generated text in scientific writing has received significant attention \cite{wu2025survey},
as has the assessment of the extent of AI use by authors \cite{bao2025there, liang2024mapping}. Wu \emph{et al.} provide a comprehensive survey of research on methods for detecting AI-generated text
\cite{wu2025survey}. They make the point that there are significant vocabulary distribution differences between AI- and human-generated texts, with AI-written texts tending to use a more limited vocabulary. They find that detection methods based on the RoBERTa LLM are quite effective \cite{liu2019roberta}. At the same time they point out that automated AI detection faces significant challenges, for instance due to out-of-distribution data or obfuscation. 

The work by Liang \emph{et al.}\cite{liang2024mapping} is particularly relevant here, as it considers  AI detection at a corpus level. They have the specific objective of estimating the proportion of texts in a corpus that are `AI-modified'. If their models are calibrated correctly, then, in 2024, 17.5\% of Computer Science papers on arXiv were AI-modified compared with just 5\% of Mathematics papers. 
A similar study considering abstracts published in PubMed in 2024 suggests that at least 13.5\% were processed using LLMs \cite{kobak2025delving}.

\subsection{Objectivity and hyperbole}
\label{sec:hype}
The over-promotion (or ``hyping") of research results by authors, and the associated dangers of  doing so, have been well-recognized in the past \cite{rinaldi2012hype,scott2017superlative}. 
More specifically, in terms of quantitative studies of the use of hyperbolic language in academic writing, there have been a number of small-scale studies based on a few hundred papers that have provided evidence of  increases in hype, in specific fields such as biomedicine \cite{fraser2009marketing,millar2019important} and sociology \cite{li2025promoting}. Later studies have validated these findings across broader sets of disciplines, e.g., \cite{hyland2021our,yuan2022academic,hyland2024hyping,qiu2024use} but again on relatively small scales (based on up to a few thousand papers or grant proposals). 

In general there has been little prior work on  large-scale quantitative analyses of hype. A notable exception is\cite{millar2023promotional} who analyzed approximately 2.4 million journal abstracts resulting from NIH-funded biomedical research from 1985 to 2020 and  found significant growth in the use of hype terms (based on a set of 139 selected hype-related adjectives). Another large-scale study is that of \cite{vinkers2015use} who analyzed 25 million PubMed abstracts from the period 1974 to 2014, finding significant increases over that time in the use of positive words relative to neutral or negative words. In this context, our work is the first that we are aware of that combines an analysis of hyperbolic language  on a large-scale (17M journal abstracts), across all academic research disciplines, in the context of the interaction between  hype and AI usage since the early 2020s.



 \subsection{Readability} \label{sec:readability}
Research on methods to quantify readability dates back to the 1920s \cite{dubay2004principles}. These methods have been applied to research literature in recent times.  Plavén-Sigray \emph{et al.} have demonstrated that readability is decreasing over time \cite{plaven2017readability}. Hengel has shown that, in economics and econometrics, papers authored by women are better written \cite{hengel2022}, with readability being quantified using five standard readability scores. The scores used are Flesch Reading Ease (FRE), Flesch-Kincaid, Gunning Fog, SMOG and Dale-Chall. These scores are representative of those used in readability analysis of research literature; the study reported in \cite{plaven2017readability} used FRE and Dale-Chall. 
These scores are inclined to be correlated as they quantify similar things. They are based on counts of polysyllabic words and sentence length (the FRE calculation is presented in section \ref{sec:meth-read}). Dale-Chall incorporates a count of `difficult words', that is words not included in a list of 3000 simple words. 


The use of tools such as FRE to analyze the readability of scientific papers has received some criticism on at least two fronts. First, it has been argued that it is completely appropriate for scientific text to use big words and have long sentences \cite{hartley2016time}, so poor scores on metrics such as the FRE should not be a concern. Secondly, surface features such as sentence length of syllable counts cannot adequately capture reading ease. Si and Callan  \cite{si2001statistical} have shown that readability scores that use statistical language models are more accurate than the Flesch-Kincaid score. Nevertheless, methods such as FRE are still widely used for assessing readability of scientific texts \cite{plaven2017readability,hengel2022}.

\section{Data, metrics, and methods}
The dataset used in this study was generated using the Semantic Scholar API during May, 2025 \cite{kinney2023semantic}. Semantic Scholar was chosen because it provides open access to a large volume of detailed publication data, without the need for costly licensing, significantly enhancing the reproducibility of this study.

\label{sec:data_methods}
\subsection{Dataset}
For this work we focused our attention on articles in the Semantic Scholar dataset which were published during the period 2000 - 2024 and whose meta-data included \emph{'JournalArticle'} in the \emph{'publicationtypes'} field. This produced an initial dataset of 29,893,335 articles, the \emph{30M} dataset.

Semantic Scholar generates its records through a combination of web crawling, partnerships with publishers, and from feeds provided directly from data providers. However, not all article records are complete. Some article records offer little more than article \emph{stubs}, based on initial information found during crawling, to be completed later as additional relevant information is located. For this study we implemented the following filtering rules to exclude outliers and to ensure we had the data (authors, references, abstracts) needed for our analysis:

\begin{enumerate}
    \item Exclude articles where abstracts had less than 5 words (35\% of articles)
    \item Exclude articles with no references (23\% of articles).
\end{enumerate}

In conjunction, these rules reduced the initial dataset by 44\%, from  29,893,335 articles to 16,970,014 articles. We refer to this smaller subset as the \emph{17M dataset} and we use it exclusively in the analysis that follows.

Each Semantic Scholar article is associated with an ordered list of \emph{fields of study} (FoS). For this work, we used the first FoS for the article's \emph{primary field of study} as a simple way to separate articles into 23 mutually exclusive sets, each corresponding to a different primary field of study.

This 17M dataset is summarised in Table \ref{tab:summary-stats}. Each row corresponds to a specific FoS, in descending order of the number of articles associated in that field. Each row shows the article count, and the mean and standard deviations of the number of authors per article (\emph{Authors}), words in the abstracts (\emph{Abs Len}), and references per article (\emph{Refs}). In addition, we indicate the percentage of abstracts (\emph{\%Abs})in the initial dataset that contained at least 5 words, since this rule was responsible for a majority of exclusions (35\%) and because it varies somewhat by FoS; for example just under 50\% of Mathematics articles passed this abstract test compared with more than 78\% of Materials Science articles.

Medicine (6,482,449 articles) and Computer Science (2,608,616 articles) account for more than half (53.5\%) of the total dataset. In contrast, Geography and Geology are the smallest sets, with just over 25,000 articles between them (only 0.15\% of the dataset). Medicine has the highest average author count (5.7±3.1 authors); Philosophy has the lowest (1.5±1.1 authors). Agriculture and Food Science (\emph{Ag/Food Sci}) have the longest abstracts (232.8±88.2 words); Chemistry has the shortest (153.1±81.2 words).

In this work, we do not make claims about the relative size of different fields of research based on these numbers. While Medicine may well be one of the largest fields of scientific research, the relative scale of Computer Science here could reflect an origin-bias in Semantic Scholar, and the relative sizes of other fields may be be greatly influenced by Semantic Scholar's data collection process.

\begin{table}
\centering

\caption{A summary of the 17M dataset used in this study showing the number of articles, the mean author and reference count, and the mean abstract length (words) per primary field of study. The table rows are sorted in descending order of the number of articles in each primary FoS group. The percentage of articles in the original 30M data that passed the abstract filter is also shown.}

\label{tab:summary-stats}

\begin{tabular}{lrllll}
\toprule
 & Articles & \%Abs & Authors & Abs Len & Refs \\
 &  &  &  &  &  \\
\midrule
Medicine & 6,482,449 & 62.8 & 5.7±3.1 & 207.8±101.5 & 39.4±36.7 \\
Computer Sci & 2,608,616 & 71.1 & 3.5±1.8 & 155.7±64.2 & 28.4±23.9 \\
Biology & 1,819,622 & 60.0 & 5.3±3.0 & 199.3±82.3 & 52.1±39.2 \\
Engineering & 1,268,123 & 74.2 & 4.2±2.2 & 165.5±69.7 & 28.3±25.4 \\
Chemistry & 927,524 & 69.8 & 5.1±2.6 & 153.1±81.2 & 44.0±37.2 \\
Env Sci & 883,431 & 57.2 & 5.3±2.9 & 211.8±82.5 & 48.2±38.6 \\
Psychology & 469,122 & 63.8 & 4.0±2.4 & 190.7±87.7 & 51.8±36.3 \\
Materials Sci & 405,632 & 78.2 & 5.6±2.8 & 174.5±62.3 & 47.1±41.5 \\
Physics & 328,824 & 79.9 & 4.3±2.8 & 148.2±66.4 & 36.9±32.7 \\
Ag/Food Sci & 266,727 & 64.5 & 5.3±2.7 & 232.8±88.2 & 43.8±33.8 \\
Education & 235,026 & 71.7 & 3.3±2.2 & 180.8±99.6 & 34.4±37.0 \\
Sociology & 211,895 & 74.3 & 3.3±2.4 & 184.0±80.8 & 46.6±29.0 \\
Mathematics & 193,256 & 49.9 & 2.5±1.3 & 129.3±74.8 & 29.3±22.2 \\
Business & 159,250 & 65.0 & 2.8±1.5 & 161.0±81.3 & 39.2±33.0 \\
Economics & 106,513 & 64.9 & 3.0±2.0 & 178.6±93.5 & 37.4±27.5 \\
Political Sci & 70,965 & 76.9 & 2.1±1.6 & 166.4±84.6 & 45.4±33.1 \\
Law & 48,289 & 64.1 & 2.5±1.9 & 173.0±112.7 & 29.0±27.8 \\
Linguistics & 45,329 & 70.5 & 2.9±1.8 & 173.7±98.3 & 42.8±35.3 \\
History & 45,077 & 68.5 & 1.9±1.8 & 176.3±120.1 & 36.7±36.7 \\
Philosophy & 26,940 & 60.1 & 1.5±1.1 & 170.9±124.7 & 33.7±36.3 \\
Art & 26,386 & 71.7 & 2.6±2.0 & 162.3±115.4 & 26.3±38.8 \\
Geography & 14,563 & 68.3 & 3.4±2.4 & 190.8±87.9 & 43.2±31.4 \\
Geology & 10,896 & 57.3 & 4.4±2.7 & 198.9±85.1 & 47.7±40.9 \\
\bottomrule
\end{tabular}
\end{table}


\subsection{Quantifying readability}\label{sec:meth-read}
To quantify readability we use the FRE score \cite{flesch1948new} discussed in section \ref{sec:readability} on the article abstracts. This score punishes long sentences with polysyllabic words so that larger scores correspond to better readability. Equation \ref{eq:FRES} shows how the FRE is calculated; $WL$ is the average word length measured in syllables and $SL$ is the average sentence length. 

\begin{equation}
    \label{eq:FRES}
    FRE = 206.835 -84.6\times WL - 1.015 \times SL
\end{equation}
The specific coefficients using in FRE are derived from a regression analysis used in the original research by Flesch \cite{flesch1948new}.

\subsection{Quantifying the strength of the AI signal}
While some research on AI tries to quantify the proportion of papers that are \emph{AI-modified}, our objective is more modest. We aim to quantify the strength of the AI signal and then examine how that varies across different fields of study, and how it correlates with hype and readability. We do this by identifying \emph{AI-amplified} terms and quantifying their frequency in document abstracts. 

This AI scoring strategy is based on a dataset of terms provided with the 2024 papers from Liang \emph{et al.} \cite{liang2024monitoring, liang2024mapping}. That dataset includes human-written and AI-generated research paper abstracts. It is important to note that these abstracts come from STEM (science, technology engineering, mathematics) disciplines so the AI score is likely to have a STEM bias. For this reason, the discussion in Section \ref{sec:discussion} concentrates on STEM FoS. 

Our AI scoring strategy assumes that AI-and human-generated text will have different vocabulary distributions. We identify the most over-represented terms in the AI-generated training data using odds ratios, defined as the ratio of the frequency of a term in the AI-generated documents relative to its frequency in human-generated documents. The set of terms with the 300 highest odds ratios are the AI-amplified terms we use in our analysis. The scoring strategy calculates the count of these terms for each document in the 17M dataset. The count is normalised by the length of the abstract text. The top 10 AI and human terms are shown in Table \ref{tab:top_terms}. The top human terms are not used in the scoring, but they are shown here for comparison purposes.


\begin{table}[h]
\caption{The top 10 AI and human terms in the training data as scored using Odds Ratio.}\label{tab:top_terms}
\begin{tabular}{l|l}
\textbf{Top AI terms} & \textbf{Top human terms} \\ \hline
delve, underscores, intricacies,  & modelled, try, nowadays, agrees, \\ 

groundbreaking, scrutinizing, meticulous, & enjoys, besides, show, equivalently, \\ 

realm, showcases, prowess, signifies &
obtains, learnt
\end{tabular}
\end{table}

While this strategy works well for papers already published, it is important to acknowledge that it may not work well in the future, as LLMs improve and usage patterns change. In fact, it seems to be an in-joke in research papers on AI detection to use `delve' in the text\cite{wu2025survey, matsui2024delving,kobak2025delving}. 

\subsection{Quantifying hype}
Research on the growth of hype in academic research typically counts the occurrence of various hype terms \cite{vinkers2015use, hyland2024hyping,millar2022trends} and that is the strategy  we follow here. We used the 139 hype terms identified by Miller \emph{et al.}\cite{millar2022trends} to generate hype scores based on the frequency of those terms, normalized by abstract length, effectively counting the number of hype terms per abstract word. Several example hype terms are shown in Table \ref{tab:hype_terms}. 


\begin{table}[h]
\caption{Sample hype terms from \cite{millar2022trends} used for hype scoring. The score is based on the frequency of occurrence of these terms normalised by abstract length. }\label{tab:hype_terms}
\begin{tabular}{c}
\textbf{Hype Terms}  \\ \hline
novel, critical, key, innovative, scientific, effective, successful, diverse, significant \\
advanced,
robust,
relevant,
strong,
unique,
comprehensive,
broad,
essential,
rigorous \\
promising,
interdisciplinary,
urgent,
quality,
first,
unmet,
outstanding,
efficient \\
\hline
\end{tabular}
\end{table}

\subsection{Analysis methods}
\label{sec:analysis_methods}
To conduct the main analysis we group the 17M dataset by year of publication and primary FoS to calculate the several key measures used in this study (see Section \ref{sec:results}, including:

\begin{itemize}
    \item The relative number of papers published, dividing the actual number of papers published in a given year/FoS by the number of papers published in 2000 for a given FoS.
    \item The mean abstract length (words) by year/FoS.
    \item The \emph{trimmed} mean of the author count by year/FoS; a $99^{th}$ percentile trimmed mean is used to deal with some fields of study (e..g. Physics) that occasionally have very large author counts, which can significantly distort the mean.
    \item The fraction of papers with a single (solo) author by year/FoS.
    \item The mean number of references per year/FoS.
    \item The mean Flesch readability score by year/FoS.
    \item The mean normalised hype term and AI term counts per year/FoS.
\end{itemize}

Regarding the mean number of authors on a paper, we also consider how the contribution per author varies with different authorship counts. For example, all other things being equal, the contribution of an author to a single-author (solo) paper should be greater than the contribution of an author to a 4-author paper. There are various ways to account for different author contributions depending on the number and position of authors, some of which are sensitive to authorship norms in different fields of study; see for example, \cite{price1963little,hagen2010harmonic,stallings2013determining,tscharntke2007author,wren2007write,lindsey2014authorship,allen2014credit}. In this work, we use the \emph{uniform fractional weighting} -- the contribution of each of the $n$ authors of a paper is estimated as $\frac{1}{n}$ -- to calculate a mean \emph{weighted output per year} for authors from the mean of the sum of their $\frac{1}{n}$ contributions in a given year. It is worth noting that, since all of the above weighting models are designed so that the contributions of authors to a paper sum to one, the mean weighted output per year does not depend on the contribution model used.

Finally, when considering AI use in papers, we focus our analysis on the post-AI period in our dataset. Specifically, we look at the relative change in the strength of the AI signal between 2020 and 2024 to identify those fields that have presented with the strongest AI signal growth over that time. We also compare the hype and readability scores for articles in the top and bottom AI signal quartiles to highlight the association that exists between AI usage and hype and readability.

\section{Results}\label{sec:results}
In this section we present the main results including a summary of our key findings by year and field of study, the results from a detailed analysis of authorship and author productivity, and the results of an analysis of AI use and its impact on objectivity and readability. Each of these results will be discussed in detail in Section \ref{sec:discussion}.

\subsection{Summary statistics}
\begin{figure}[t!]
  \centering
\includegraphics[width=\textwidth]{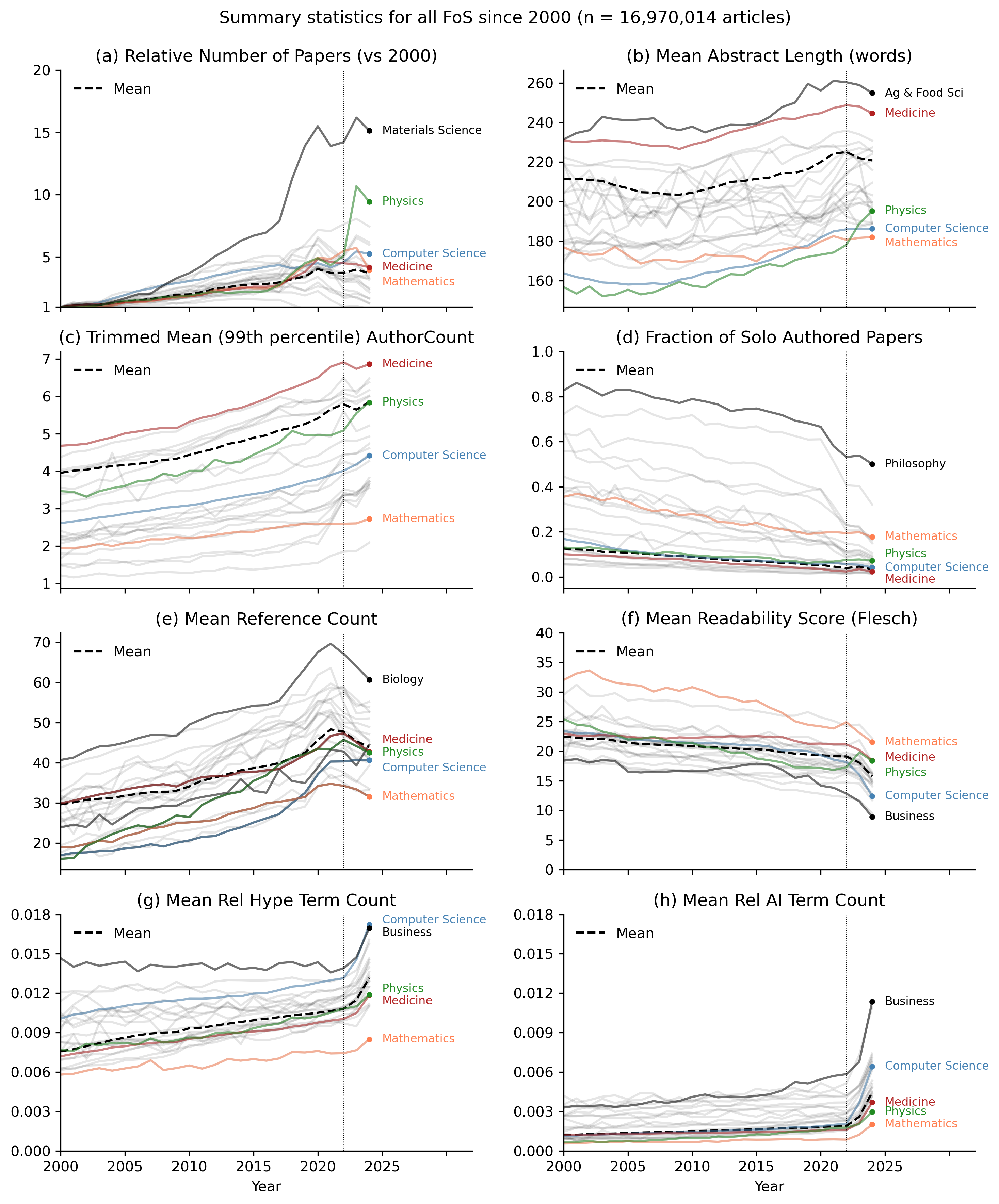}
  \caption{Summary results for all FoS since the year 2000. In each graph, the dashed line shows the mean value across all FoS.}
  \label{fig:main_stats}
\end{figure}

Figure \ref{fig:main_stats} summarises the main results using the metrics from Section \ref{sec:analysis_methods} by publication year and FoS. Each graph shows the trend lines for each FoS during 2000 - 2024, with a dashed line to indicate the aggregate mean across all fields of study. In each graph, several fields of study of particular interest are highlighted and labeled. 

In advance of a more detailed discussion of these results in Section \ref{sec:discussion} it is worth making the following iniitial observations:

\begin{itemize}
    \item There has been a threefold increase in the relative number of papers per year between 2000 and 2024; see Figure \ref{fig:main_stats}(a). This includes a very significant increase in Physics papers, increasing from 6,451 in 2000 to 49,201 in 2024. This result is not surprising and similar observations have been reported elsewhere \cite{hanson2024strain}.
    \item Paper abstracts and bibliographies have been getting longer (Figure \ref{fig:main_stats}(b) and (e), respectively), particularly since 2010, as more and more papers became available electronically and page-counts were relaxed. 
    \item Different FoS have very different author counts; see Figure \ref{fig:main_stats}(c). Medical papers now have more than six authors on average while Mathematics papers have fewer than three. There has been a steady increase in author count across all fields and, not surprisingly, the proportion of solo authored papers has been decreasing; Figure \ref{fig:main_stats}(d). 
    \item Readability, as quantified by the Flesch Reading Ease score (long sentences and polysyllabic words), has been declining -- Figure \ref{fig:main_stats}(f) -- and this decline accelerates after 2023, probably as a result of the uptake of AI tools. 
    \item At the same time we see an increase in the use of hype terms (see Table \ref{tab:hype_terms}) in Figure \ref{fig:main_stats}(g). There is a marked increase from 2023, again presumably linked with the uptake of AI. 
    \item Finally, in the post-AI era (2023 and later) there has been a striking increase in the strength of the AI score indicating that this metric is effective at picking up an AI signal; Fig \ref{fig:main_stats}(h). 
\end{itemize}


\subsection{Author count}
Figure \ref{fig:main_stats}(c)
shows that the author count has increased for all FoS since 2000. For example, for Computer Science, the average number of authors per paper has increased from 2.6 in 2000 to 4.6 in 2024. 

\subsubsection{Author count and multidisciplinarity}
In an effort to understand this increase in author count, we looked in detail at five well-known Computer Science/Engineering journals that are more or less likely to attract research that is multidisciplinary in nature. For this particular analysis we used a secondary dataset made available in \cite{cunningham2024analysis} focused on Computer Science publications. For example, \emph{IEEE Transactions on Information Theory} and \emph{Artificial Intelligence} would normally publish single discipline papers, whereas the \emph{Journal of the American Medical Informatics Association (JAMIA)}, \emph{AI in Medicine} and \emph{Computational Biology and Chemistry} tend to attract many multidisciplinary papers. 

\begin{figure}[t!]
  \centering
\includegraphics[width=0.7\textwidth]{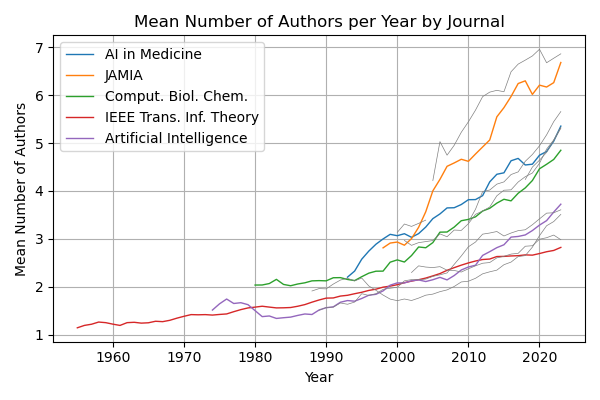}
  \caption{A closer look at authorship statistics for several Computer Science/Engineering journals with varying degrees of multidisciplinary research. 
  } 
  \label{fig:CS_venues_authors}
\end{figure}

The mean authors counts for papers appearing in these journals are shown in Figure \ref{fig:CS_venues_authors} for several decades. Later in Section \ref{sec:discussion} we will consider these results as a possible explanation for increasing author counts.

\subsubsection{Weighted author count and productivity}
\label{sec:authorprod}
\begin{figure}[t!]
  \centering
\includegraphics[width=1\textwidth]{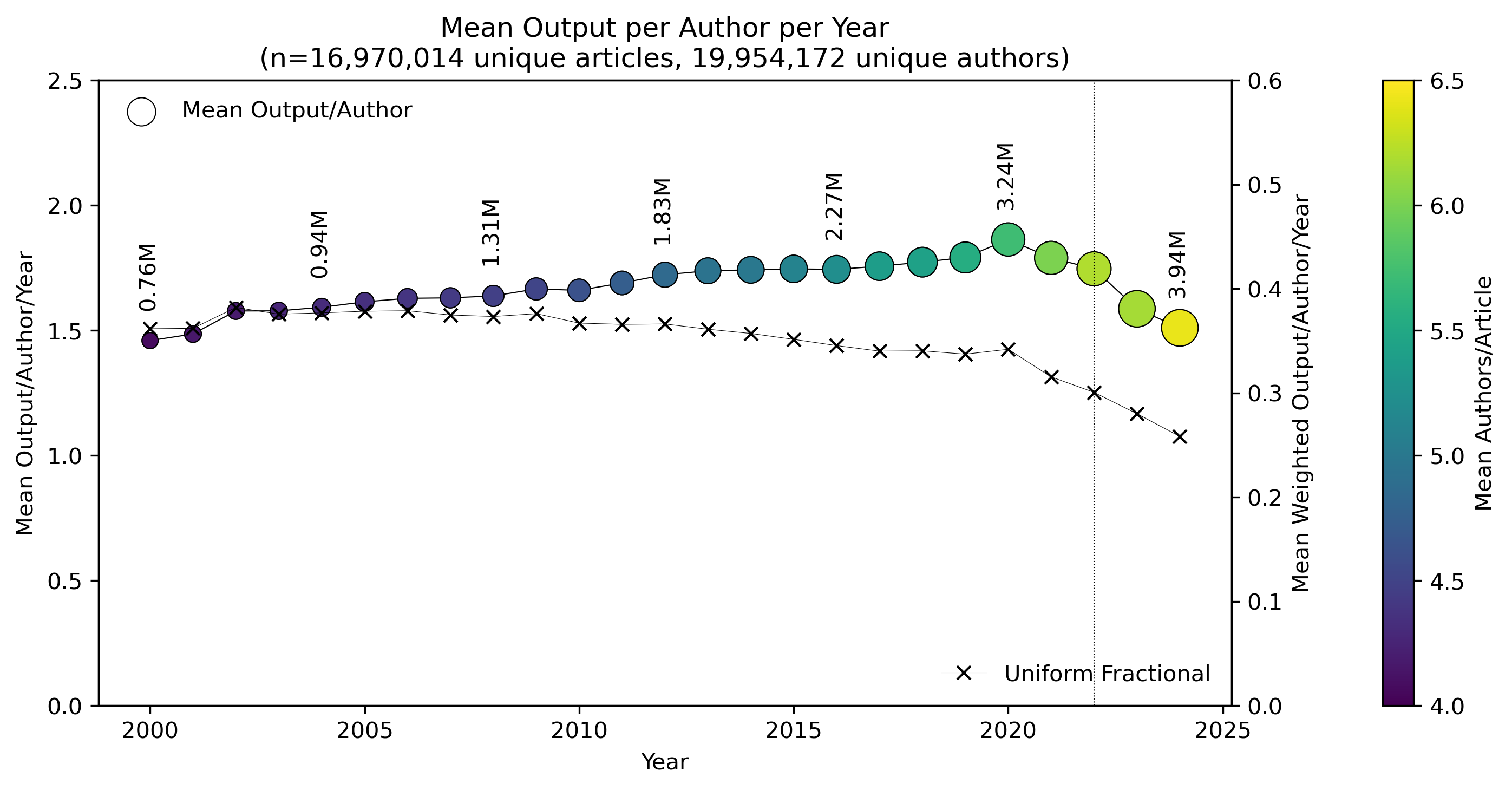}
  \caption{Changing authorship and estimating author productivity. The line with the colour-coded markers shows the mean number of papers published per year per author across all fields of study, and indicates that author output has been steadily increasing up to 2020. The size of these markers indicates the total number of articles in our dataset that year, with several years labeled for reference. The colour-coding indicates the average number of authors per article, which has also been steadily increasing. The second (dashed) line is an estimate of author productivity based on the sum of the \emph{weighted} number of articles per author per year. In this case an author on an article with $n$ authors receives a weighted output of $1/n$ for that article (\emph{uniform fractional} weighting).}
  \label{fig:AuthorProd}
\end{figure}

Given the significant increase in papers published per year (Figure \ref{fig:main_stats}(a)) and these changes in authorship, it is also worth considering the relationship between author counts, relative author contributions, and overall author productivity. The colour-coded line in Figure \ref{fig:AuthorProd} indicates the mean number of papers published per year per author across all fields of study. This line shows that author output has been steadily increasing, at least up to 2020; the reason for the decline in recent years appears to be an artifact of the Semantic Scholar data corresponding to a natural latency that exists in the Semantic Scholar data collection process; we shall discuss this further in Section \ref{sec:limitations}.

\begin{figure}[t!]
  \centering
\includegraphics[width=1\textwidth]{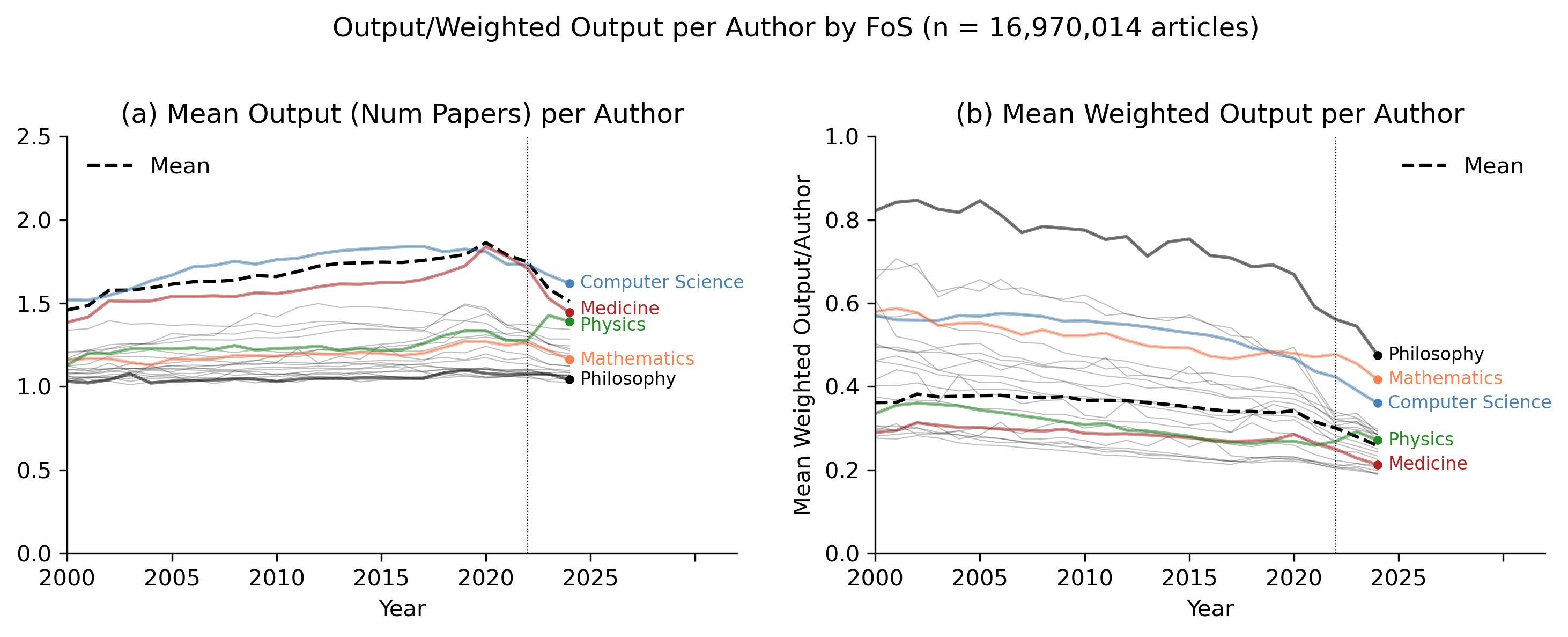}
  \caption{Estimating author productivity by field of study: (a) the mean output per author per year for each field of study; (b) the corresponding weighted output per author per year, as an indicator of author productivity, for each field of study using the uniform fractional weighting model.}
  \label{fig:AuthorProdFoS}
\end{figure}

The second (dashed) line in Figure \ref{fig:AuthorProd} shows the mean weighted output per author per year (as an indicator of \emph{weighted productivity}) using the \emph{uniform fractional weighting} model; each author of a paper with $n$ authors, receives a weight of $\frac{1}{n}$. As previously mentioned, due to the \emph{sum-to-one} nature of different authorship weighting models, this approach is independent of any particular authorship weighting model. Figure \ref{fig:AuthorProdFoS} shows these productivity results by FoS and we will discuss the difference between these unweighted and weighted productivity results further in Section \ref{sec:discussion}, and why they might indicate undesirable authorship practices such as gift authorship.

\subsection{AI, hype, and readability}
\label{sec:AI_res} 
In addition to changing authorship practices we are also interested in how content practices have been evolving, particularly in recent years with the appearance of AI tools.

\subsubsection{AI signal strength}
Figure \ref{fig:AI_change} looks at the change in the AI signal (estimated by a normalised count of AI-amplified terms) has changed between 2020 and 2024. In Figure \ref{fig:AI_change}(a), each FoS is positioned according to the mean AI score of its papers in 2020 (x-axis) and 2024 (y-axis). Each FoS is indicated by a colour-coded marker. Marker size corresponds to the total number of articles in the dataset for that FoS and marker colour correspond to the percentage change in AI score between 2020 and 2024; each FoS is also labeled with this percentage change.

\subsubsection{Hype/readability and AI use}
\begin{figure}[t]
  \centering
\includegraphics[width=\textwidth]{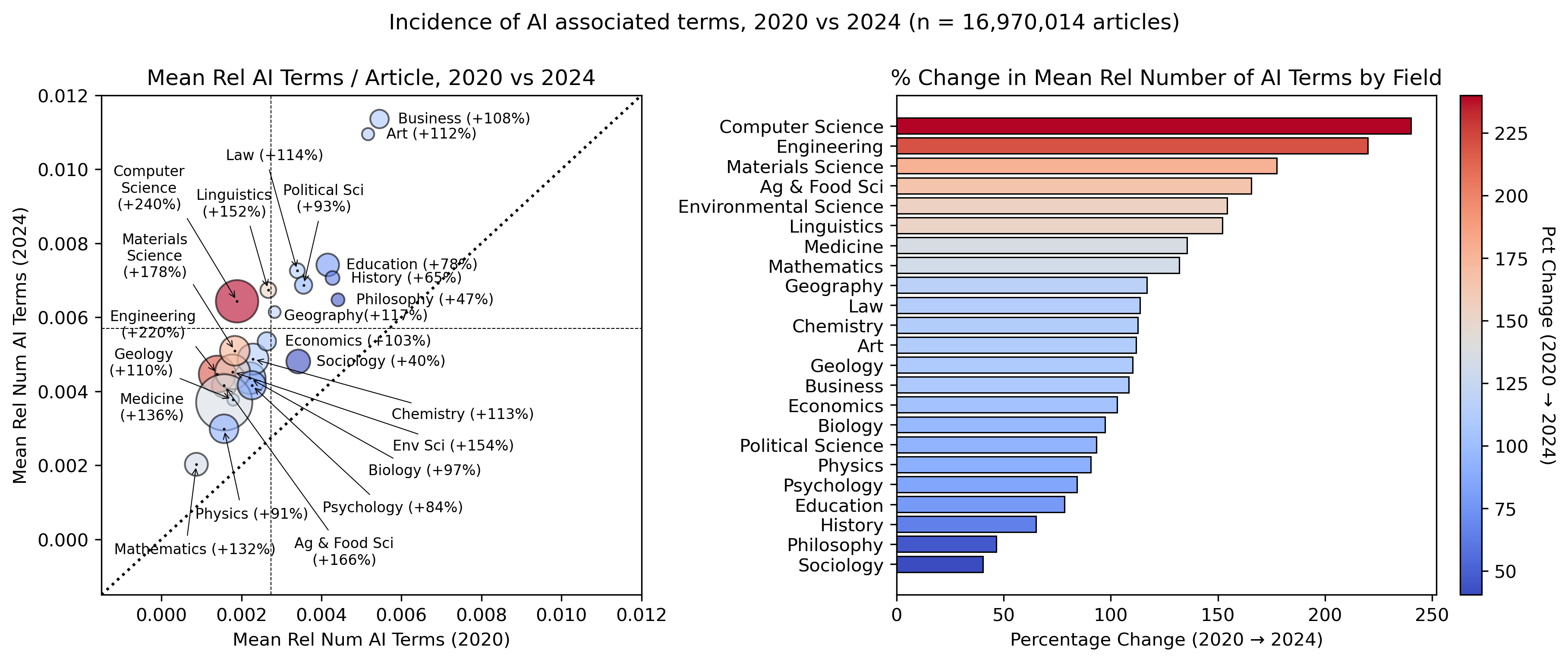}
  \caption{The increase in the incidence of AI-associated terms across all FoS between 2020 and 2024. The dotted line on the left indicates an equal incidence between 2020 and 2024. All FoS show an increased incidence.  }
  \label{fig:AI_change}
\end{figure}

Figure \ref{fig:AI_change}(b) ranks the fields of study according to this percentage change in AI score to show which fields have been more or less impacted by AI use. 




\begin{figure}[ht]
  \centering
\includegraphics[width=\textwidth]{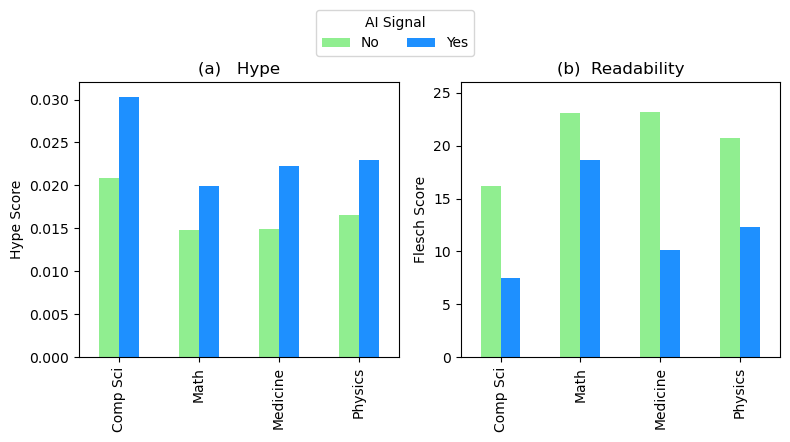}
  \caption{An analysis of differences between the top quartile and bottom quartile in terms of AI Signal across four FoS selected from the ranking in Figure \ref{fig:AI_change}. The blue bars represent the top quartile indicating papers where the use of AI is most likely.} 
  \label{fig:likely_AI}
\end{figure}

The frequency of AI-associated terms can be used to rank papers by the likelihood of AI use. To explore this we distinguish between papers (in the post-AI period) that present with and without an AI signal, using the top and bottom quartiles of papers by AI score, respectively. The results in Figure \ref{fig:likely_AI} compare the (a) hype and (b) readability scores for these groups in four STEM fields --- Computer Science, Mathematics, Medicine and Physics -- chosen to reflect different degrees of AI usage change based on Figure \ref{fig:AI_change}(b);  Computer Science is at the top of the ranking, Physics is towards the bottom, while Medicine and Mathematics are closer to the middle. 

\section{Discussion}
\label{sec:discussion}
The results presented above show several trends in various authorship and content practices and, in particular, on the growing use of AI in academic publishing. In this section we discuss these findings in more detail, their possible causes, and their wider impact. 

\subsection{Authorship practices}
While the increase in the number of authors on research papers has been commented on elsewhere \cite{hosseini2022ethical,jakab2024many,cunningham2024analysis,reisig2020assessing}, we are not aware of any research that quantifies how pervasive this is across different fields. 
Figure \ref{fig:main_stats}(a) shows that the number of papers being produced is rising rapidly overall, and across many different fields of study. This increase has been contributed to significantly by the emergence of new OA publishers \cite{cunningham2024analysis,shen2015predatory} and it adds a significant burden on the peer review process resulting in poorer oversight and lower quality papers. 

\subsubsection{What is driving author inflation?}
It is clear in Figure \ref{fig:main_stats}(c) that the mean number of authors per paper is increasing across all fields of study. This inflationary phenomenon has been discussed elsewhere \cite{jakab2024many,cunningham2024analysis,brand2015beyond} with a variety of explanations proposed, including:
\begin{itemize}
    \item research groups are larger due to increased research funding;
    \item there is more multidisciplinary research, drawing in researchers from several disciplines;
    \item improved communication is leading to more collaborations between institutions and with industry;
    \item inappropriate behaviors such as gift authorship or paper mills \cite{brand2015beyond}.
\end{itemize}

Most likely a combination of these factors are at play and in the following subsections we consider this in more detail.

\subsubsection{Increasing multidisciplinary research}
The expectation that multidisciplinary papers  have more authors seems to hold true. The results in Figure \ref{fig:CS_venues_authors}, show that multidisciplinary journals, such as the \emph{Journal of the American Medical Informatics Association (JAMIA)}, \emph{AI in Medicine}, and \emph{Computational Biology and Chemistry}, exhibit higher author counts than single-discipline journals, such as \emph{IEEE Transactions on Information Theory} and \emph{Artificial Intelligence}. However, the author counts for all five journals have been rising steadily since 1990, so an increase in multidisciplinary research, on its own, does not explain the overall increase in author count. 

\subsubsection{Author inflation and declining productivity}
One plausible explanation is that the observed increases may be due to more relaxed authorship practices over time \cite{hosseini2022ethical,jakab2024many,cunningham2024analysis,reisig2020assessing}, and an increase in so-called 'gift' authorship \cite{zauner2018we} or related poor practices \cite{brand2015beyond}; see section \ref{sec:changineauthorcounts}. Identifying instances of gift authorship is challenging, but the declining productivity signal in Figures \ref{fig:AuthorProd} and \ref{fig:AuthorProdFoS} may help to estimate the degree of the problem. 

We might reasonably expect increasing author counts to translate into a corresponding increase in publication output. If we assume an author has a finite amount of effort to invest in publications, then this effort can be focused on a small number of low-authorship papers, where the author invests a greater amount of effort per paper, or spread more thinly across a larger number of high-authorship papers, where less per paper effort is needed. Thus, authors should invest a similar total effort per year under different authorship regimes. However, Figures \ref{fig:AuthorProd} and \ref{fig:AuthorProdFoS} show that as mean author count has increased, mean weighted output (total weighted effort) has declined. In other words, when authors are involved in papers with large numbers of co-authors, the overall the sum of these contributions is less that we might expect, all other things being equal.

Of course, all other things are not equal. The total effort required for multi-author, and especially multidisciplinary, papers is likely to be greater than for single author papers or papers with lower author counts. Multi-author/multidisciplinary papers will carry a greater \emph{collaboration cost}: meetings need to be arranged and coordinated; effort needs to be parceled out; individual authors need to familiarise themselves with the norms of unfamiliar disciplines etc. Indeed, such costs may be disproportionately greater in certain fields. For example, clinical research in Medicine and related fields may be associated with a greater cost, involving more specialities, and more onerous research controls, than research in pure Mathematics.

However, at the same time, there have been significant advances in the technology and tools available to support this such collaborative, multidisciplinary endeavours: the internet has greatly reduced the cost of collaboration; shared research infrastructure and modern software tools have enhanced our ability to perform research at scale;  new tools have emerged to support collaborative writing etc. These advances should reduce the cost of collaboration, and yet this is not evident overall (see Figure \ref{fig:AuthorProd}) or in individual fields of study (see Figure \ref{fig:AuthorProdFoS}(b)).

One potential explanation for this could be that these productivity gains have been nullified by poor behaviours, such as gift authorship. If one of the authors on a 4-author paper is not materially contributing to the effort, this will dilute the relative contributions of the other 3 authors -- each will receive a weighted contribution of 0.25 instead of the more deserved contribution of 0.33 -- which will depress the weighted output estimates. In Figure \ref{fig:AuthorProd}, the weighted output of authors per year has decreased by about 10\% over the period 2000-2020 and by more than 20\% in the last 5 years. If we assume that increases in collaboration costs are balanced by recent productivity gains, then the continued decline in weighted output might suggest that gift authorship, or related practices, are becoming more and more prevalent. Indeed, it has been suggested in other studies on specific research areas (neuroscience and medicine) that gift authorship is having a significant impact on authorship inflation \cite{an2020authorship,lin2023exponential}.




\subsubsection{Dealing with author inflation}
In response, it has been argued that clarity on the requirements for authorship \cite{zauner2018we} and explicit declarations about contribution \cite{brand2015beyond} will help to address such authorship issues. However, it seems clear that these measures are not working, at least not on their own. We also need to de-incentivize the unwarranted inclusion of authors on papers. One practical way to do this is to modify metrics such as the $h$-index and field-weighted citation impact (FWCI) to account for author counts. The idea that citation metrics should be normalised to consider author count is not new \cite{koltun2021h}, but it is increasingly clear that it is necessary. To be blunt: impact assessments without some form of authorship normalisation are no longer valid.

\subsection{Content practices}
Content practices have been changing alongside authorship practices. Once again, these changes are not always positive -- declining objectivity and readability -- and there is evidence that they have accelerated in the post-AI era.

\subsubsection{Decreasing objectivity, increasing hyperbole}

The research presented here extends earlier work \cite{hyland2024hyping,vinkers2015use} discussed in section \ref{sec:hype} in two important ways: (i) It shows that this phenomenon is pervasive, being evident across the 23 FoS in the Semantic Scholar corpus (see Figure \ref{fig:main_stats}(g)); and (ii) it shows that this increase is happening right up to the present and even accelerating in recent years, perhaps with the arrival of AI tools in since late 2022.

Figure \ref{fig:main_stats}(g) shows an increasing trend in the average number of hype terms per abstract word, overall and across the 23 individual fields of study. For most of the period studied the Business field exhibits the greatest hype-term use, but Computer Science has caught up, especially in the post-AI years, and now ranks as number one in hype-term usage. Other fields, such as Mathematics, are less inclined to use hype-terms, but even Mathematics has registered significant hype-term growth in the post-AI era.

\subsubsection{Decreasing readability}
Another disappointing finding, relating to the content of research papers, is a consistent decline in readability. The mean Flesch Reading Ease Score has been decreasing gradually, across all fields of study, as shown in Figure \ref{fig:AI_change}(f). This score penalises long sentences and polysylabic words, and while it can be argued that it is appropriate for research papers to contain long sentences and big words, it is still the case that papers have been becoming more and more difficult to read. 

As was the case with hype, the field of Business is associated with poor readability scores for most of the period studied, with an accelerated decline since 2015. In contrast, and perhaps surprisingly, Mathematics papers are associated with some of the best readability scores during the same period. Care should be taken to not read too much into the readability findings between different fields, since certain fields may be biased towards lower or higher FRE scores because of the nature of their content (complex words, Greek symbols, even equations). Greater emphasis can be placed on changes in FRE within a fields; see \cite{Yeung2018_readability_neuroimaging, Balz2022_remote_sensing_readability,Paz2009_health_quality_of_life_readability}

Furthermore, just as the post-AI era is associated with an increase in hype-term usage, so too we see an accelerated decline in readability during this time period in Figure \ref{fig:main_stats}(f). For example, the readability of Computer Science and Business papers falls by more than 25\% in 2024 relative to 2020 levels and the average drop in readability across FoS is just over 18\% between 2020 and 2024. In the next section, we will look in more detail at these post-AI era changes to determine the likelihood that they are due to increased use of AI tools.

\subsection{On the impact of AI}
Figure \ref{fig:main_stats}(h) shows the proportion of AI-amplified terms in abstracts; this is presented as a population-level measure of AI signal strength rather than an article-based metric. While these terms have been in evidence throughout the analysis period, they are more or less common in different fields of study, but their frequency has been relatively stable. However, it is clear that since 2023 there has been a noteworthy increase in frequency, indicating that this measure is sensitive to increasing AI usage. 

\subsubsection{Pre vs post-AI periods}
Looking in more detail at this increase in Figure \ref{fig:AI_change}, which compares this AI signal strength between 2020 and 2024, Computer Science and Engineering exhibit the greatest increase in this measure of AI signal strength. In both cases, the frequency of AI-associated terms has increased by more than 200\% in 2024, compared to 2020. In contrast, fields such History, Philosophy, and Sociology are associated with a much more modest, but still significant, increase of 50-60\%. It may be that, since the AI terms may have a STEM bias, the detection on non-STEM papers is not as effective. Nevertheless, the picture for STEM subjects can be considered reliable and, for instance, the difference between Computer Science and Physics is probably real. 

Admittedly in this study we are only able to consider the content of abstracts, but it is reasonable to assume that if such tools are being used to help produce abstracts, then they are also likely to be used throughout the paper-writing process. It is also not so surprising to see that computer scientists are the most enthusiastic adopters of such technologies.

\subsubsection{With \& without AI signal}
If AI tools are being used more, then what does this mean for academic papers? Aside from the issues of ethics and fairness, surrounding AI use, is there any evidence that AI usage is changing the content of the papers produced? 

The results in Figure \ref{fig:likely_AI} shed some light on this question by comparing papers with and without an AI signal, in terms of their (a) hype and (b) readability scores, in four STEM fields. The average hype scores are significantly higher ($\approx30\%$ for papers with an AI signal (top quartile by AI score) compared to those without an AI signal (bottom quartile by AI score), regardless of STEM FoS. Similarly, the readability scores are lower ($\approx25-50\%$) for papers with an AI signal, compared to those without an AI signal. This is clear evidence that the strength of the AI signal correlates with an increased use of hype terms -- a tendency to sensationalize the presentation of the research or a loss of objectivity -- and a decline in readability, because of longer sentences and more polysyllabic words.


Given the number of papers involved in the analysis, it is not surprising that these results are all statistically significant. When the differences in means are analyzed with a two-tailed $t$-test, all the $p$-values are effectively zero. The effect sizes were also analyzed using Cohen's d statistic\cite{cohen2013statistical}. This is a simple measure that considers the differences in means divided by the pooled standard deviation of the data. For the Hype Scores, these vary between 0.31 for Mathematics and 0.51 for Medicine. A score of 0.5 would be considered a `Medium' effect size. For the Flesch reading ease scores the Cohen's d statistic varies between 0.16 for Mathematics to 0.76 for Medicine. This would be considered a `Large` effect size for Medicine.




\subsection{Limitations}\label{sec:limitations}
There are several limitations of this work that should be considered when evaluating its findings. In this section we consider these in detail, paying particular attention to limitations associated with the dataset used in this study, as well as several considerations when it comes to the measurement of hype and AI signals.

\subsubsection{Dataset limitations}
This work uses an open-source dataset of publication records and meta-data made available by Sematic Scholar. This is in contrast with alternative sources of data, e.g. Web of Science or Scopus, used in other scientometric studies \cite{baghini2024usability,haruna2025innovation,zhu2020tale}. Still, Semantic Scholar is an established dataset for scientometrics research \cite{cunningham2024analysis,he2025academic,10.1145/3742442,smyth2019tale,liang2021finding} and it is well-suited for the present study. It is also freely available, which offers significant benefits in terms of the reproducibility of this work; Web of Science and Scopus datasets are only available under license.

Notwithstanding Semantic Scholar's open source benefits, it has several limitations worth noting. The data can be noisy. For example, although we have focused on articles that are tagged as journal articles this does not guarantee that all of the articles are journal articles or that all journal articles have been included. Similarly the 'field of study' meta-data includes a mixture of categories that are externally sourced or automatically generated, or both, which are likely to be imperfect. 

Data records can also be incomplete. In this work, for example, we began with a dataset of almost 30M records we ended up using a subset of just under 17M records, after dropping incomplete records, which were missing abstracts or reference lists. There is also some evidence in our results of a further gap in Semantic Scholar records in recent years. For example, notice in Figure \ref{fig:main_stats}(a) how the mean relative number of papers appears to flatten after 2020 and, notice too, there is a drop in the average number of papers per author in Figure \ref{fig:AuthorProd} during the same time period. This is likely due to missing data records as a result of Semantic Scholar's data collection process. While many papers are ingested through regular and reliable feeds from publishers and other sources, Semantic Scholar also collects data by crawling reference links to produce new record stubs, which may take time to flesh out into complete records. 

In summary, Semantic Scholar is a good fit for this study. It offers important reproducibilty benefits. But, like most datasets it is imperfect. That being said, we believe that its shortcomings do not compromise results presented in this work.

\subsubsection{Hype term limitations}
In this work we have used one particular set of hype terms to quantify hype term usage in abstracts: the 139-term set proposed by \cite{millar2022trends}. This is not a definitive hype-term list as other lists of similar terms do exist; see for example, \cite{hyland2024hyping} and \cite{vinkers2015use}. However, it is worth noting that we have found our findings to be largely unchanged when these alternative hype-term lists are used. For example, we produced a merged list of the 55 terms from Hyland and Jiang \cite{hyland2024hyping} and the 25 terms from Vinkers \emph{et al.}\cite{vinkers2015use} and found the results produced to be very similar to those presented here. 

\subsubsection{Detecting an AI signal}
Our AI scoring strategy is based on a dataset provided with the 2024 papers from Liang \emph{et al.} \cite{liang2024monitoring, liang2024mapping} which propose a population-level statistical framework, with models trained on just under one million papers during 2021-2024 from several fields (Computer Science, Electrical Engineering and Systems Science, Mathematics, Physics, and Statistics), to estimate the degree of AI-modified content in these fields over time.

Although we used a similar set of AI terms, we did not adopt this pre-trained framework, which begs the question whether or not our AI signal correlates closely with the predictions from this alternative approach. To test this we applied these pre-trained models to predict the prevalence of AI content in a subset of fields from our Semantic Scholar dataset as proxies for the fields used in \cite{liang2024mapping}: Computer Science, Engineering, Mathematics, Physics, and Biology. In each case, the predictions obtained for the period 2021-2024 correlated closely with the predictions presented in \cite{liang2024mapping} with $r^2>0.98$ for each FoS tested. Thus, we can conclude that our approach to AI detection is closely aligned with that used by \cite{liang2024monitoring,liang2024mapping}, at least in a subset of comparable fields.

It also is worth noting that this analysis has been conducted over a much larger dataset  and many more fields of study (almost 17M papers, 23 fields). However, as mentioned previously, the AI-associated terms used to generate our AI score were derived from the more limited STEM corpora used by \cite{liang2024monitoring,liang2024mapping}. Thus, the AI signal for non-STEM FoS may be less reliable than that for STEM FoS; it may have a STEM bias. This is one reason why our later analyses focused on STEM fields.

Finally, while the use of AI in scientific writing is likely to remain, at least the negative impact, in terms of hype and readability, may not be so bad in the future. New AI tools may be able to do better in these terms. It is also important to emphasise that the AI scoring mechanism used here will probably not work well in the future as the vocabulary distributions of AI tools move closer to that of humans. 

\section{Conclusion}
The operation of the overall research publication system incentivizes more and more publications in an environment where the main currency is papers and citations. It has long been understood that this has a negative impact on publication quality \cite{fire2019over}. The findings of this paper, based on an analysis of 17M papers published since 2000 in 23 fields of study, show that this is indeed the case. 

The main results in our study are that as publication volume has increased significantly, author counts have increased and readability has declined. Furthermore, there is a clear move towards hyperbolic claims about research significance. There is evidence of the use of AI is paper writing since the emergence of commercial LLMs in late 2022; Computer Science and Engineering stand out in this regard. And, where there is evidence of the use of AI, it seems to have a negative impact on hype and readability. 

It is difficult to incentivize researchers to produce fewer, higher quality papers in this environment. However, it should be possible to keep the number of authors in check and reduce poor authorship practices. Currently, publication metrics such as the $h$-index and FWCI do not normalize by author count. Normalized versions of these metrics have been proposed \cite{bihari2023review,pepe2012measure} and if they were more widely adopted, by academic institutions, funding agencies, and publishers, it would surely dis-incentivize the practice of gifting authorship to fellow researchers.  

Finally, it is clear that the mainstream availability of AI tools is having a major impact. Since these tools can be a great help to non-native English speakers, their use should be accepted in some circumstances. For the moment, it is essential that publishers have a policy whereby all use of AI tools in paper writing is transparent to readers and explicitly declared. It is also vital that researchers continue to have oversight and responsibility for the content of their research papers.  

\section*{Acknowledgments}
Supported by Science Foundation Ireland through the Insight Centre for Data Analytics (12/RC/2289\_P2). PS was supported by the Hasso Plattner Institute (HPI) Research Center in Machine Learning and Data Science at the University of California, Irvine and by gift funding from SAP. 


%
%
%

\begin{thebibliography}{10}

\bibitem{10.1145/3097983.3098016}
Dong Y, Ma H, Shen Z, Wang K.
\newblock A Century of Science: Globalization of Scientific Collaborations, Citations, and Innovations.
\newblock In: Proceedings of the 23rd ACM SIGKDD International Conference on Knowledge Discovery and Data Mining. KDD '17. New York, NY, USA: Association for Computing Machinery; 2017. p. 1437–1446.
\newblock Available from: \url{https://doi.org/10.1145/3097983.3098016}.

\bibitem{doi:10.1128/mbio.02515-24}
Casadevall A, Clark LF, Fang FC.
\newblock The changing roles of scientific journals.
\newblock mBio. 2024;15(11):e02515--24.
\newblock doi:{10.1128/mbio.02515-24}.

\bibitem{cunningham2024analysis}
Cunningham P, Smyth B.
\newblock An Analysis of the Impact of Gold Open Access Publications in Computer Science.
\newblock Commun ACM. 2025;68(9).
\newblock doi:{10.1145/3721975}.

\bibitem{osborne2019authorship}
Osborne JW, Holland A.
\newblock What is authorship, and what should it be? A survey of prominent guidelines for determining authorship in scientific publications.
\newblock Practical Assessment, Research, and Evaluation. 2019;14(1):15.

\bibitem{mills2021problematizing}
Mills D, Inouye K.
\newblock Problematizing ‘predatory publishing’: A systematic review of factors shaping publishing motives, decisions, and experiences.
\newblock Learned Publishing. 2021;34(2):89--104.

\bibitem{severin2021overburdening}
Severin A, Chataway J.
\newblock Overburdening of peer reviewers: A multi-stakeholder perspective on causes and effects.
\newblock Learned publishing. 2021;34(4):537--546.

\bibitem{hanson2024strain}
Hanson MA, Barreiro PG, Crosetto P, Brockington D.
\newblock The strain on scientific publishing.
\newblock Quantitative Science Studies. 2024;5(4):823--843.

\bibitem{jakab2024many}
Jakab M, Kittl E, Kiesslich T.
\newblock How many authors are (too) many? A retrospective, descriptive analysis of authorship in biomedical publications.
\newblock Scientometrics. 2024;129(3):1299--1328.

\bibitem{hosseini2022ethical}
Hosseini M, Lewis J, Zwart H, Gordijn B.
\newblock An ethical exploration of increased average number of authors per publication.
\newblock Science and Engineering Ethics. 2022;28(3):25.

\bibitem{flanagin1998prevalence}
Flanagin A, Carey LA, Fontanarosa PB, Phillips SG, Pace BP, Lundberg GD, et~al.
\newblock Prevalence of articles with honorary authors and ghost authors in peer-reviewed medical journals.
\newblock Jama. 1998;280(3):222--224.

\bibitem{reisig2020assessing}
Reisig MD, Holtfreter K, Berzofsky ME.
\newblock Assessing the perceived prevalence of research fraud among faculty at research-intensive universities in the USA.
\newblock Accountability in Research. 2020;27(7):457--475.

\bibitem{brand2015beyond}
Brand A, Allen L, Altman M, Hlava M, Scott J.
\newblock Beyond authorship: Attribution, contribution, collaboration, and credit.
\newblock Learned Publishing. 2015;28(2).

\bibitem{zauner2018we}
Zauner H, Nogoy NA, Edmunds SC, Zhou H, Goodman L. We need to talk about authorship; 2018.

\bibitem{wu2025survey}
Wu J, Yang S, Zhan R, Yuan Y, Chao LS, Wong DF.
\newblock A survey on llm-generated text detection: Necessity, methods, and future directions.
\newblock Computational Linguistics. 2025;51(1):275--338.

\bibitem{bao2025there}
Bao H, Sun M, Teplitskiy M.
\newblock Where there’sa will there’sa way: ChatGPT is used more for science in countries where it is prohibited.
\newblock Quantitative Science Studies. 2025; p. 1--23.

\bibitem{liang2024mapping}
Liang W, Zhang Y, Wu Z, Lepp H, Ji W, Zhao X, et~al.
\newblock Mapping the increasing use of LLMs in scientific papers.
\newblock arXiv preprint arXiv:240401268. 2024;.

\bibitem{liu2019roberta}
Liu Y, Ott M, Goyal N, Du J, Joshi M, Chen D, et~al.
\newblock Roberta: A robustly optimized bert pretraining approach.
\newblock arXiv preprint arXiv:190711692. 2019;.

\bibitem{kobak2025delving}
Kobak D, Gonz{\'a}lez-M{\'a}rquez R, Horv{\'a}t E{\'A}, Lause J.
\newblock Delving into LLM-assisted writing in biomedical publications through excess vocabulary.
\newblock Science Advances. 2025;11(27):eadt3813.

\bibitem{rinaldi2012hype}
Rinaldi A.
\newblock To hype, or not to (o) hype: communication of science is often tarnished by sensationalization, for which both scientists and the media are responsible.
\newblock EMBO reports. 2012;13(4):303--307.

\bibitem{scott2017superlative}
Scott SL, Jones CW. Superlative scientific writing; 2017.

\bibitem{fraser2009marketing}
Fraser VJ, Martin JG.
\newblock Marketing data: Has the rise of impact factor led to the fall of objective language in the scientific article?
\newblock Respiratory Research. 2009;10(1):35.

\bibitem{millar2019important}
Millar N, Salager-Meyer F, Budgell B.
\newblock “It is important to reinforce the importance of…”:‘Hype’in reports of randomized controlled trials.
\newblock English for Specific Purposes. 2019;54:139--151.

\bibitem{li2025promoting}
Li Z, Lin J, Xu J.
\newblock Promoting research: Academic hypes embedded in the rhetorical move structure of sociology research article abstracts.
\newblock Journal of English for Academic Purposes. 2025;76:101535.

\bibitem{hyland2021our}
Hyland K, Jiang FK.
\newblock ‘Our striking results demonstrate…’: Persuasion and the growth of academic hype.
\newblock Journal of Pragmatics. 2021;182:189--202.

\bibitem{yuan2022academic}
Yuan Zm, Yao M.
\newblock Is academic writing becoming more positive? A large-scale diachronic case study of {S}cience research articles across 25 years.
\newblock Scientometrics. 2022;127(11):6191--6207.

\bibitem{hyland2024hyping}
Hyland K, Jiang F.
\newblock Hyping the REF: promotional elements in impact submissions.
\newblock Higher Education. 2024;87(3):685--702.

\bibitem{qiu2024use}
Qiu HS, Peng H, Fosse HB, Woodruff TK, Uzzi B.
\newblock Use of promotional language in grant applications and grant success.
\newblock JAMA Network Open. 2024;7(12):e2448696--e2448696.

\bibitem{millar2023promotional}
Millar N, Batalo B, Budgell B.
\newblock Promotional Language (hype) in abstracts of publications of {N}ational {I}nstitutes of {H}ealth--funded research, 1985-2020.
\newblock JAMA Network Open. 2023;6(12):e2348706--e2348706.

\bibitem{vinkers2015use}
Vinkers CH, Tijdink JK, Otte WM.
\newblock Use of positive and negative words in scientific PubMed abstracts between 1974 and 2014: retrospective analysis.
\newblock BMJ. 2015;351.

\bibitem{dubay2004principles}
DuBay WH.
\newblock The principles of readability.
\newblock Online submission. 2004;.

\bibitem{plaven2017readability}
Plav{\'e}n-Sigray P, Matheson GJ, Schiffler BC, Thompson WH.
\newblock The readability of scientific texts is decreasing over time.
\newblock Elife. 2017;6:e27725.

\bibitem{hengel2022}
Hengel E.
\newblock Publishing While Female: are Women Held to Higher Standards? Evidence from Peer Review.
\newblock The Economic Journal. 2022;132(648):2951--2991.
\newblock doi:{10.1093/ej/ueac032}.

\bibitem{hartley2016time}
Hartley J.
\newblock Is time up for the Flesch measure of reading ease?
\newblock Scientometrics. 2016;107(3):1523--1526.

\bibitem{si2001statistical}
Si L, Callan J.
\newblock A statistical model for scientific readability.
\newblock In: Proceedings of the tenth international conference on Information and knowledge management; 2001. p. 574--576.

\bibitem{kinney2023semantic}
Kinney R, Anastasiades C, Authur R, Beltagy I, Bragg J, Buraczynski A, et~al.
\newblock The semantic scholar open data platform.
\newblock arXiv preprint arXiv:230110140. 2023;.

\bibitem{flesch1948new}
Flesch R.
\newblock A new readability yardstick.
\newblock Journal of applied psychology. 1948;32(3):221.

\bibitem{liang2024monitoring}
Liang W, Izzo Z, Zhang Y, Lepp H, Cao H, Zhao X, et~al.
\newblock Monitoring ai-modified content at scale: A case study on the impact of chatgpt on ai conference peer reviews.
\newblock arXiv preprint arXiv:240307183. 2024;.

\bibitem{matsui2024delving}
Matsui K.
\newblock Delving into PubMed records: Some terms in medical writing have drastically changed after the arrival of ChatGPT.
\newblock MedRxiv. 2024; p. 2024--05.

\bibitem{millar2022trends}
Millar N, Batalo B, Budgell B.
\newblock Trends in the use of promotional language (hype) in abstracts of successful national institutes of health grant applications, 1985-2020.
\newblock JAMA network open. 2022;5(8):e2228676--e2228676.

\bibitem{price1963little}
Price DJdS.
\newblock Little Science, Big Science.
\newblock New York: Columbia University Press; 1963.

\bibitem{hagen2010harmonic}
Hagen NT.
\newblock Harmonic allocation of authorship credit: Source-level correction of bibliometric bias assures accurate publication and citation analysis.
\newblock PLOS ONE. 2010;5(12):e12348.
\newblock doi:{10.1371/journal.pone.0012348}.

\bibitem{stallings2013determining}
Stallings J, Vance S, Yang J, Vannier M, Liang J, Pang L, et~al.
\newblock Determining scientific impact using a collaboration index.
\newblock Scientometrics. 2013;95(2):597--610.
\newblock doi:{10.1007/s11192-012-0824-4}.

\bibitem{tscharntke2007author}
Tscharntke T, Hochberg ME, Rand TA, Resh VH, Krauss J.
\newblock Author sequence and credit for contributions in multiauthored publications.
\newblock PLoS Biology. 2007;5(1):e18.
\newblock doi:{10.1371/journal.pbio.0050018}.

\bibitem{wren2007write}
Wren JD, Kozak KZ, Johnson KR, Deakyne SJ, Schilling LM, Dellavalle RP.
\newblock The write position: A survey of perceived contributions to papers based on byline position and number of authors.
\newblock EMBO reports. 2007;8(11):988--991.
\newblock doi:{10.1038/sj.embor.7401093}.

\bibitem{lindsey2014authorship}
Lindsey MA, Yager J.
\newblock Authorship and corresponding author issues in nursing research publications: Ethical considerations.
\newblock Nursing Ethics. 2014;21(2):169--178.
\newblock doi:{10.1177/0969733013508979}.

\bibitem{allen2014credit}
Allen L, Scott J, Brand A, Hlava M, Altman M.
\newblock Credit where credit is due.
\newblock Nature. 2014;508(7496):312--313.
\newblock doi:{10.1038/508312a}.

\bibitem{shen2015predatory}
Shen C, Bj{\"o}rk BC.
\newblock ‘Predatory’open access: a longitudinal study of article volumes and market characteristics.
\newblock BMC medicine. 2015;13(1):230.

\bibitem{an2020authorship}
An JY, Marchalik RJ, Sherrer RL, Baiocco JA, Rais-Bahrami S.
\newblock Authorship growth in contemporary medical literature.
\newblock SAGE Open Medicine. 2020;8:2050312120915399.

\bibitem{lin2023exponential}
Lin Z, Lu S.
\newblock Exponential authorship inflation in neuroscience and psychology from the 1950s to the 2020s.
\newblock American Psychologist. 2023;.

\bibitem{koltun2021h}
Koltun V, Hafner D.
\newblock The h-index is no longer an effective correlate of scientific reputation.
\newblock PLoS One. 2021;16(6):e0253397.

\bibitem{Yeung2018_readability_neuroimaging}
Yeung AWK, Goto TK, Leung WK.
\newblock Readability of the 100 Most‐Cited Neuroimaging Papers Assessed by Common Readability Formulae.
\newblock Frontiers in Human Neuroscience. 2018;12:308.
\newblock doi:{10.3389/fnhum.2018.00308}.

\bibitem{Balz2022_remote_sensing_readability}
Balz T, et~al.
\newblock Scientometric Full-Text Analysis of Papers Published in Remote Sensing.
\newblock Remote Sensing. 2022;14(17).
\newblock doi:{10.3390/rs14174285}.

\bibitem{Paz2009_health_quality_of_life_readability}
Paz SH, et~al.
\newblock Readability estimates for commonly used health‐related quality‐of‐life instruments.
\newblock Quality of Life Research. 2009;18(8):889--900.
\newblock doi:{10.1007/s11136-009-9504-1}.

\bibitem{cohen2013statistical}
Cohen J.
\newblock Statistical power analysis for the behavioral sciences.
\newblock Routledge; 2013.

\bibitem{baghini2024usability}
Baghini MS, Mohammadi M, Norouzkhani N.
\newblock Usability Testing: A Bibliometric Analysis Based on WoS Data.
\newblock Journal of Scientometric Research. 2024;13(1):9--24.
\newblock doi:{10.5530/jscires.13.1.2}.

\bibitem{haruna2025innovation}
Haruna EU, Asiedu WK, Baek YJ.
\newblock Mapping the Research Trends on Technological Innovation in East Asia: A Bibliometric Analysis Using the Scopus Database for Future Research Direction (1982--2022).
\newblock Journal of Scientometric Research. 2025;13(3s):s3--21.
\newblock doi:{10.5530/jscires.20041153}.

\bibitem{zhu2020tale}
Zhu J, Liu W.
\newblock A tale of two databases: the use of Web of Science and Scopus in academic papers.
\newblock Scientometrics. 2020;123:321--335.
\newblock doi:{10.1007/s11192-020-03387-8}.

\bibitem{he2025academic}
He G, Yuan J, Yang Y.
\newblock Measuring Academic Cocoon from Disparity and Diversity Perspectives.
\newblock Scientometrics. 2025;130:2445--2474.
\newblock doi:{10.1007/s11192-025-05290-6}.

\bibitem{10.1145/3742442}
Smyth B.
\newblock People Who Liked This Also Liked ... A Publication Analysis of Three Decades of Recommender Systems Research.
\newblock ACM Trans Recomm Syst. 2025;doi:{10.1145/3742442}.

\bibitem{smyth2019tale}
Smyth B.
\newblock A Tale of Two Communities: An Analysis of Three Decades of Case-Based Reasoning Research.
\newblock In: Bach K, Marling C, editors. Case-Based Reasoning Research and Development: 27th International Conference, ICCBR 2019, Otzenhausen, Germany, September 8–12, 2019, Proceedings. vol. 11680 of Lecture Notes in Computer Science. Springer, Cham; 2019. p. 343--357.

\bibitem{liang2021finding}
Liang Z, Mao J, Lu K, Li G.
\newblock Finding citations for PubMed: a large-scale comparison between five freely available bibliographic data sources.
\newblock Scientometrics. 2021;126(12):9519--9542.
\newblock doi:{10.1007/s11192-021-04191-8}.

\bibitem{fire2019over}
Fire M, Guestrin C.
\newblock Over-optimization of academic publishing metrics: observing Goodhart's Law in action.
\newblock GigaScience. 2019;8(6):giz053.

\bibitem{bihari2023review}
Bihari A, Tripathi S, Deepak A.
\newblock A review on h-index and its alternative indices.
\newblock Journal of Information Science. 2023;49(3):624--665.

\bibitem{pepe2012measure}
Pepe A, Kurtz MJ.
\newblock A measure of total research impact independent of time and discipline.
\newblock PLoS One. 2012;7(11):e46428.

\end{thebibliography}

\end{document}